\documentclass[12pt]{article}

\pdfoutput=1

\usepackage{amsmath,amssymb}
\usepackage{epsfig}
\usepackage{color}
\usepackage{verbatim}
\usepackage[normalem]{ulem}
\usepackage[utf8]{inputenc}

\begin{document}

\begin{titlepage}

\title{\bf\Large Detecting the brightest HAWC sources with IceCube \\
in the upcoming years}
\author{
Viviana Niro \thanks{Email: \tt viviana.niro@apc.in2p3.fr} \\ \\
{\normalsize \it Université de Paris, CNRS, Astroparticule et Cosmologie, F-75006 Paris, France}\\
}

\date{\today}
\maketitle

\thispagestyle{empty}

\begin{abstract}
Among the gamma-ray sources detected by the HAWC observatory, 
we consider in details in this work the gamma-ray sources eHWC J1907+063 and 
eHWC J2019+368. 
These two sources belong to the three most luminous sources detected by HAWC, with emission above 100~TeV. 
In addition to those, we have considered also a source for which IceCube currently 
report an excess in neutrinos, the 2HWC J1857+027 source. 
For these sources, we show the prediction for neutrinos at the IceCube detector. 
Moreover, we present the calculation of the statistical significance, 
considering 10 and 20 years of running time, and we comment on the 
current results reported by the collaboration. We found that a 
detection at 3$\sigma$ or more should be within reach of the next decade 
for the sources eHWC J1907+063 and 
eHWC J2019+368. Instead, the detection at about 3$\sigma$ of the 2HWC J1857+027 
source will depend on the specific value of the flux, on the extension of the source and on the cut-off 
energy.  
\end{abstract}

\end{titlepage}

\section{\label{sec:intro} Introduction}
Neutrinos are particles that rarely interact with matter and that are unaffected by
magnetic fields. 
Therefore, they can travel undeflected through cosmological distances, 
providing important information on some of the most energetic and distant phenomena of the Universe.
They can shed light on the origin of cosmic-rays (CR) and 
the gamma-ray emission. 
Through their detection, it is, indeed, possible to discriminate between leptonic and hadronic 
particle acceleration scenarios. In the leptonic scenario, gamma-rays are produced through
processes like bremsstrahlung, synchrotron radiation and inverse Compton scattering. In the hadronic
scenario, instead, gamma-rays are produced from the decay of neutral pions. In the latter case, 
from the decay of charged pions, also neutrinos are produced. For this 
reason, neutrino telescopes can unambiguously probe the hadronic particle acceleration scenario.
The identification of the origin of the gamma-ray emission, specifically if it is leptonic 
or hadronic is one of the most important goals in gamma-ray astronomy~\cite{Felix}. 

The IceCube detector has reported 
103 neutrino events of astrophysical origin, of which 60 events with deposited energy 
$E_{dep}>$ 60 TeV~\cite{Schneider:2019ayi}, considering 7.5 years of running time. 
The current event distribution is consistent with isotropy. For this reason, it is 
often interpreted in terms of extragalactic sources, see e.g. 
Ref.~\cite{Ahlers:2018dtq}. 

Several studies have been carried out previously about the possible detection of Galactic sources 
in the northern hemisphere at IceCube, in particular considering sources detected by the Milagro Collaboration, 
among which MGRO J1908+063 and MGRO J2019+368, see e.g.~\cite{Halzen:2008zj,Kappes:2009zza,
GonzalezGarcia:2009jc,Gonzalez-Garcia:2013iha,Halzen:2016seh}. 

In a previous study, Ref.~\cite{Gonzalez-Garcia:2013iha}, the authors revisited the prospects for observing the Milagro sources 
in light of the low-energy cut-off reported by the Milagro collaboration~\cite{Abdo:2012jg,Smith:2010yn}. 
Subsequently, in Ref.~\cite{Halzen:2016seh}, it was concluded that for MGRO J1908+06 
an evidence at $3\sigma$ could be obtained in about ten years assuming values 
of the spectral index and the cut-off energy that are in good agreement with the best fit 
reported in~\cite{Aharonian:2009je}. 
The answer depends on the neutrino energy threshold considered in the specific analysis. 
In general, however, in about 15 years of IceCube data the sources MGRO J1908+06 
and MGRO J2019+37 should be detectable. 

The HAWC observatory has reported new data on galactic sources in recent years, see 
e.g.~\cite{Abeysekara:2017hyn,Malone:2018zeg,Abeysekara:2019gov}. 
In the 2HWC catalog~\cite{Abeysekara:2017hyn}, 39 gamma-ray sources 
were identified, with best sensitivity at about 7 TeV energies. The fit was done 
using a power-law spectrum, without an energy cut-off and considering two hypotheses 
for the sources: a point-source case and an extended emission within an uniform disk 
of fixed radius. 
An error of about 50\% on the flux normalization was reported and an error of 0.1$^\circ$ on 
the tested radius. 
Recently, all HAWC sources present in the 2HWC catalogue have been considered, 
excluding those for which the flux can fully be ascribed to a pulsar wind nebulae (PWN)
\footnote{Note that recently the IceCube collaboration has also presented a search for neutrinos coming 
from PWN~\cite{Aartsen:2020eof}.} and 
analysed in comparison with the IceCube data~\cite{Kheirandish:2019bke}. 
Different analyses have been considered for the HAWC sources: 
sources in the northern hemisphere, the Cygnus region, and in particular the 
2HWC J1908+063 and 2HWC J1857+027 region~\cite{Kheirandish:2019bke}, reporting 
a p-value of about 2\% from the 2HWC J1857+027 region. 

Subsequently, in the eHWC catalogue of Ref.~\cite{Abeysekara:2019gov}, nine sources were 
observed above 56~TeV, all of which are likely Galactic in origin. 
Among these sources, the eHWC J1825-134 source, located in the southern sky, has been detected with an 
hard spectrum that extends up to multi-TeV energies, 
thus it represents a possible PeVatron source. 
Moreover, this is the brightest source detected by HAWC in the multi-TeV domain. 
For this reason, it was analysed in details in Ref.~\cite{Niro:2019mzw}, 
specifically considering prediction for the KM3NeT detector 
and the possibility of discovering the source at the Baikal-GVD experiment and at the 
IceCube detector. 

The sources eHWC J1907+063 and eHWC J2019+368 were identified with the 
eHWC catalogue as well~\cite{Abeysekara:2019gov}, as two of the three gamma-ray sources, 
together with the eHWC J1825-134 source, that emits above 100~TeV. Other six sources 
were identified above 56~TeV. The source 2HWC J1857+027 does not belong to 
this list. Note, however, that IceCube 
has reported a p-value of about 2\% from this source~\cite{Kheirandish:2019bke}. Moreover, for 
six of the sources that emits above 56~TeV, an 
integrated flux was reported in the eHWC catalogue, assuming an 
$E^{-2}$ or $E^{-2.7}$ spectrum. Some of the sources 
detected by HAWC, that are in the eHWC catalogue, were previously detected also by Milagro. 

The paper is organized as the following. In Section~\ref{sec:HAWC}, we describe the data 
reported on these sources, eHWC J1907+063, eHWC J2019+368 (2HWC J2019+367) and 2HWC J1857+027, by the HAWC collaboration, while in 
Section~\ref{sec:nu_flux}, the calculation of the neutrino events from the sources and 
the atmospheric background is considered. 
In Section~\ref{sec:res}, we present our results on the p-value analysis and 
Section~\ref{sec:conclusion} contains our conclusions.

\section{\label{sec:HAWC} HAWC sources}

The new information from gamma-ray experiments turns out to be important for a better parametrization of the flux of the gamma-ray sources. 
The uncertainties in the normalization, spectrum and extension of the sources can result in important variations in the prediction of the neutrino fluxes. 
In this context, using updated data is important to make more reliable predictions and more correct interpretations of the 
IceCube results.

Within the eHWC catalogue~\cite{Abeysekara:2019edl,Abeysekara:2019gov}, it was found that a better fit to the gamma-ray spectrum of 
individual sources is given by a log parabola, instead of a power law: 
\begin{eqnarray}
\frac{dN_\gamma}{dE_\gamma} &=& \phi_0~\left(\frac{E_\gamma}{10~{\rm TeV}}\right)^{-\alpha_\gamma-\beta~{\rm ln}(E/{\rm 10~TeV})}
\label{eq:ehwc}
\end{eqnarray}
with $\alpha_\gamma$ the spectral index and $\phi_0$ the flux normalization, 
see values in Table~\ref{tab:sources_fit} for the sources eHWC J1907+063 and eHWC J2019+368. 
The source eHWC J1825-134 was, instead, studied in details in~\cite{Niro:2019mzw} 
in connection to the KM3NeT detector.   

The explicit values of the systematic errors on the normalization of the flux vary with 
energy and are reported in~\cite{Abeysekara:2019gov}. They are of the order of the one reported for the Crab 
nebula~\cite{Abeysekara:2019edl}, i.e. of the order of about 15\%. To simplify our analysis we did not 
considered them in the following. 

The emission reported in the 2HWC catalogue, instead, is given for point-sources and for 
a bin radius and it is parametrized as: 
\begin{eqnarray}
\frac{dN_\gamma}{dE_\gamma} &=& \phi_0~\left(\frac{E_\gamma}{7~{\rm TeV}}\right)^{-\alpha_\gamma}
\label{eq:2hwc}
\end{eqnarray}
where no cut-off is being reported. We report in Table~\ref{tab:sources_fit_2hwc}, the parameters for the 
sources 2HWC J1908+063, 2HWC J2019+367, and 2HWC J1857+027 for the case of extended emission. 
For the values of the parameters in the case of point-source hypothesis, we refer to 
Ref.~\cite{Abeysekara:2017hyn}. 
In the left panels of Fig.~\ref{fig:gamma_flux_first}, Fig.~\ref{fig:gamma_flux_second} and~\ref{fig:gamma_flux_third} 
are reported the spectra given in the eHWC and 2HWC catalogue.

\begin{table}[!t]
\centering
\begin{tabular}{l ||c| c | c | c | c  }
\hline
Source & Dec($^\circ$)& $\sigma_{ext}(^\circ)$ & $\phi_0$  & $\alpha_\gamma$ & $\beta$ \\ \hline \hline
eHWC J1907+063 & $6.32 \pm 0.09$ & $0.67\pm0.03$ & $0.95\pm0.05$ & $2.46\pm 0.03$ & $0.11\pm0.02$ \\[0.5ex]
eHWC J2019+368 & $36.78 \pm 0.04$ & $0.30\pm0.02$ & $0.45\pm0.03$ & $2.08\pm 0.06$ & $0.26\pm0.05$ \\[0.5ex] \hline
\end{tabular}
\caption{\label{tab:sources_fit}Declination, extension of the source in degrees, 
normalization of the flux $\phi_0$ in units of $10^{-13}~{\rm TeV}^{-1}~{\rm cm}^{-2}~{\rm s}^{-1}$, spectral index $\alpha_\gamma$ and 
$\beta$ parameter, see Eq.~\ref{eq:ehwc}, for 
two of the most luminous sources in the eHWC catalogue, eHWC J1907+063 and 
eHWC J2019+368. 
}
\end{table}

\begin{table}[!t]
\centering
\begin{tabular}{l ||c| c | c | c }
\hline
Source & Dec($^\circ$)& $r_{bin}(^\circ)$ & $\phi_0$  & $\alpha_\gamma$ \\ \hline \hline
2HWC J1908+063 & $6.39$ & $0.8$ & $85.1\pm 4.2$ & $-2.33\pm 0.03$  \\[0.5ex]
2HWC J2019+367 & $36.80$ & $0.7$ & $58.2\pm4.6$ & $-2.24\pm 0.04$ \\[0.5ex] 
2HWC J1857+027 & 2.80 & 0.9 & $97.3 \pm 4.4$ & $-2.61\pm 0,04$ \\\hline
\end{tabular}
\caption{\label{tab:sources_fit_2hwc}Declination, tested radius in degrees, 
normalization of the flux $\phi_0$ in units of $10^{-15}~{\rm TeV}^{-1}~{\rm cm}^{-2}~{\rm s}^{-1}$ and spectral index $\alpha_\gamma$, 
see Eq.~\ref{eq:2hwc}. Besides, the sources 2HWC J1908+063 and 2HWC J2019+367, 
we have also considered the source 2HWC J1857+027, for which an excess in neutrinos has been reported, 
see text for more details. 
}
\end{table}

The 2HWC J1908+063 source has been studied in the recent source analysis search done by the IceCube 
collaboration, that has reported a p-value of about 1\% from this source, see Ref.~\cite{Aartsen:2018ywr}. 
This source, initially dubbed MGRO~J1908+06, was first detected by the Milagro experiment~\cite{Abdo:2007ad,Abdo:2009ku,Smith:2010yn} 
and subsequently by the ARGO-YBJ experiment~\cite{ARGO-YBJ:2012goa}. 
It has been then detected by HESS~\cite{Aharonian:2009je} and 
VERITAS~\cite{Aliu:2014xra}. 
It was then detected by HAWC~\cite{Abeysekara:2015qba} in 2015.  
As it was noted in Ref.~\cite{Halzen:2016seh}, even if the Fermi-LAT observes a pulsar within the extension 
of the source~\cite{Abdo:2010ht}, the large size of the source is maybe consistent with a supernova remnant, 
that are the more accredited sources of the highest energy cosmic rays in the Galaxy~\cite{BaadeAndZwicky}, see also 
Ref.~\cite{Ackermann:2013wqa,Gabici:2007qb}~\footnote{Note that the Fermi Large Area Telescope has detected 
gamma-ray spectra of the supernova remnants IC 443 and W44 that are compatible with a pion-decay feature~\cite{Ackermann:2013wqa}}. 
While studying this source, we will consider only the latest parametrization as reported in the 
eHWC catalogue~\cite{Abeysekara:2019gov}.

The 2HWC J2019+367 source belongs to the Cygnus region, that is a complex region of about~5$^\circ$ where 
five 2HWC sources can be found, of which one is most probably associated with the Cygnus Cocoon 
field~\cite{Abeysekara:2017hyn}. 
This source has been detected in the 2HWC catalogue within a radius of 
0.7$^\circ$. The emission from this region is compatible with the one detected previously 
by the Milagro experiments as coming from MGRO J2019+37, see 
Refs.~\cite{Abdo:2007ad,Abdo:2009ku,Abdo:2012jg}. VERITAS~\cite{Aliu:2014rha} has resolved the 
emission into two sources, 
VER J2016+371 and the brighter VER J2019+368. The first emission could be associated 
with the supernova remnant CTB 87 or a blazar, while the second could be associated 
with two pulsars and a star-forming region, see Ref.~\cite{Abeysekara:2017hyn} for 
a detailed description about the different emission components. 

Considering the complexity of the Cygnus region, we have 
decided to consider for this source the extended emission from a radius of 0.7$^\circ$, as 
reported in the 2HWC catalogue, as well as the parametrization for 
eHWC J2019+368 reported in the eHWC catalogue~\cite{Abeysekara:2019gov}. 
We will always work under the assumption that 
the $\gamma-$ray production from this region 
is hadronic and thus the neutrino flux will be calculated using 
the standard formalism highlighted in Sect.~\ref{sec:nu_flux}. 

The source 2HWC J1857+027 has been classified in the 2HWC catalogue and studied in 
connection with the IceCube data in Ref~\cite{Kheirandish:2019bke}. 
In that analysis an excess of neutrinos was found from the 2HWC J1857+027 region, resulting 
in a p-value of about 2\%. Concerning gamma-ray subsequent studies carried out by the 
HAWC collaboration, namely the eHWC catalogue, this source failed to pass the threshold 
to belong to the sources with high-energy emission, explicitly above 56~TeV. 
Note, instead, that the eHWC J1850+001 source belongs to the eHWC catalogue, 
but this source is at a declination of $0.14 \pm 0.12$. 
This means that the emission from the 2HWC~J1857+027 source above 56~TeV is fainter than the emission 
from the sources present in the eHWC catalogue. 
Nevertheless, we have considered this source since an excess in neutrinos is present. 
For this reason, we have compared the integrated flux above 56~TeV from this source with the 
ones reported for the others eHWC sources. Fixing $\alpha_\gamma$ to the best-fit value and 
considering a systematic error of +50\% in the normalization, we find the integrated flux 
to be about $8.32 \times 10^{-15}~{\rm~cm^{-2}~s^{-1}}$ for $E_{cut, \gamma}$=100~TeV, while considering the normalization 
best-fit we find $5.55 \times 10^{-15}~{\rm~cm^{-2}~s^{-1}}$. 
Considering $E_{cut, \gamma}$=300~TeV, instead, we find integrated fluxes  
of about $1.5 \times 10^{-14}~{\rm~cm^{-2}~s^{-1}}$ and 
$10^{-14}~{\rm~cm^{-2}~s^{-1}}$, respectively. 
On the other hand, the fainter source of the eHWC catalogue has an integrated 
flux of about $(0.9\pm 0.2) \times 10^{-14}~{\rm ph~cm^{-2}~s^{-1}}$. 
We will show in the following how the systematic error on the normalization of the source 2HWC J1857+027 
has an impact on the statistical significance. 
Moreover, as exemplification, we will consider as cut-off energy 50~TeV, 100~TeV and 300~TeV. 
In the left panel of Fig.~\ref{fig:gamma_flux_third}, we reported the resulted spectrum and, 
as comparison, we show also the spectrum for the most luminous source of the 
eHWC catalogue, eHWC J1825-134, that has a power law with exponential cut-off fit.

\section{\label{sec:nu_flux} Neutrino event rate}

\begin{figure}[!h]
\centering
\begin{tabular}{rl}
\includegraphics[width=0.48\textwidth]{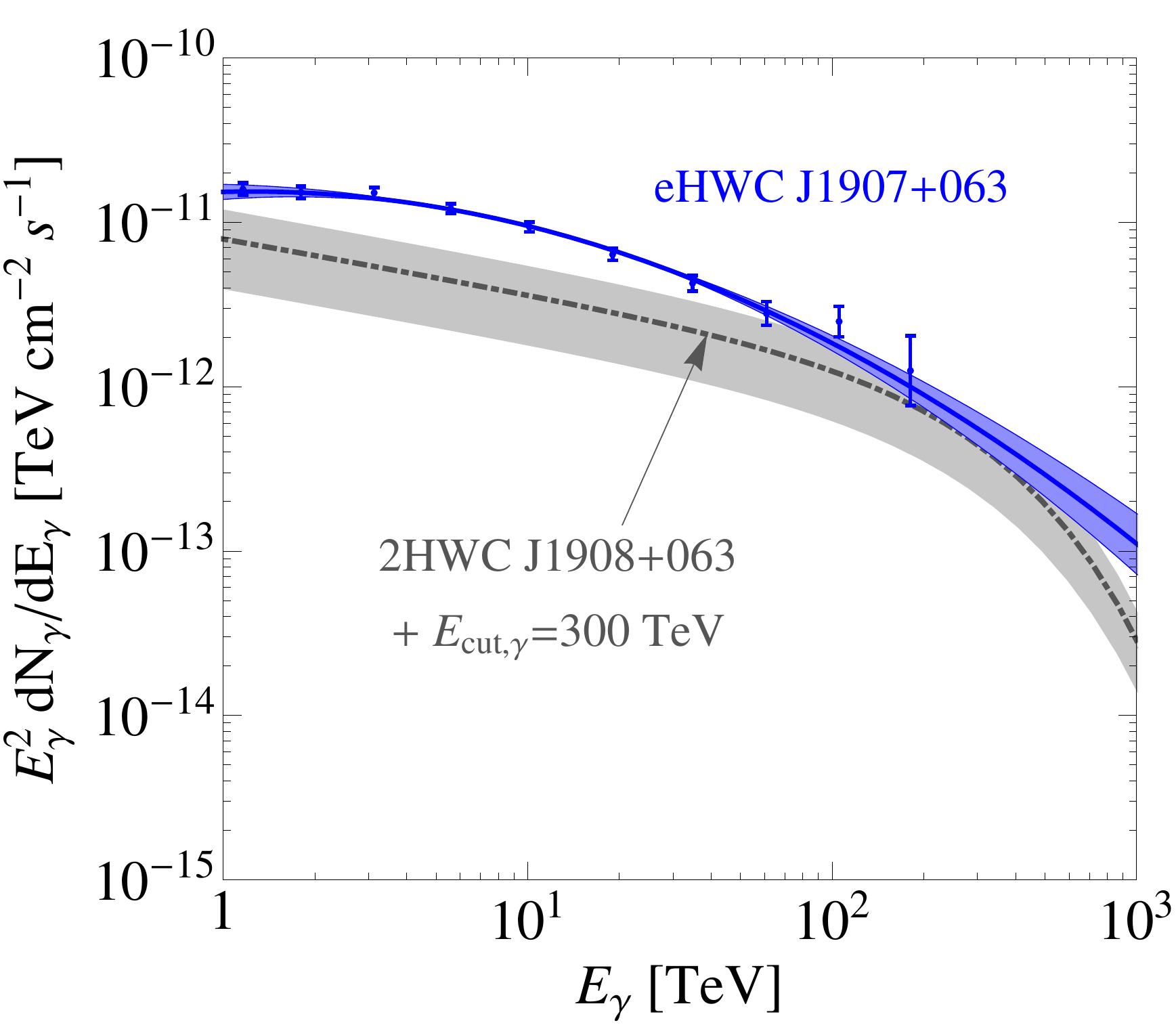} &
\includegraphics[width=0.48\textwidth]{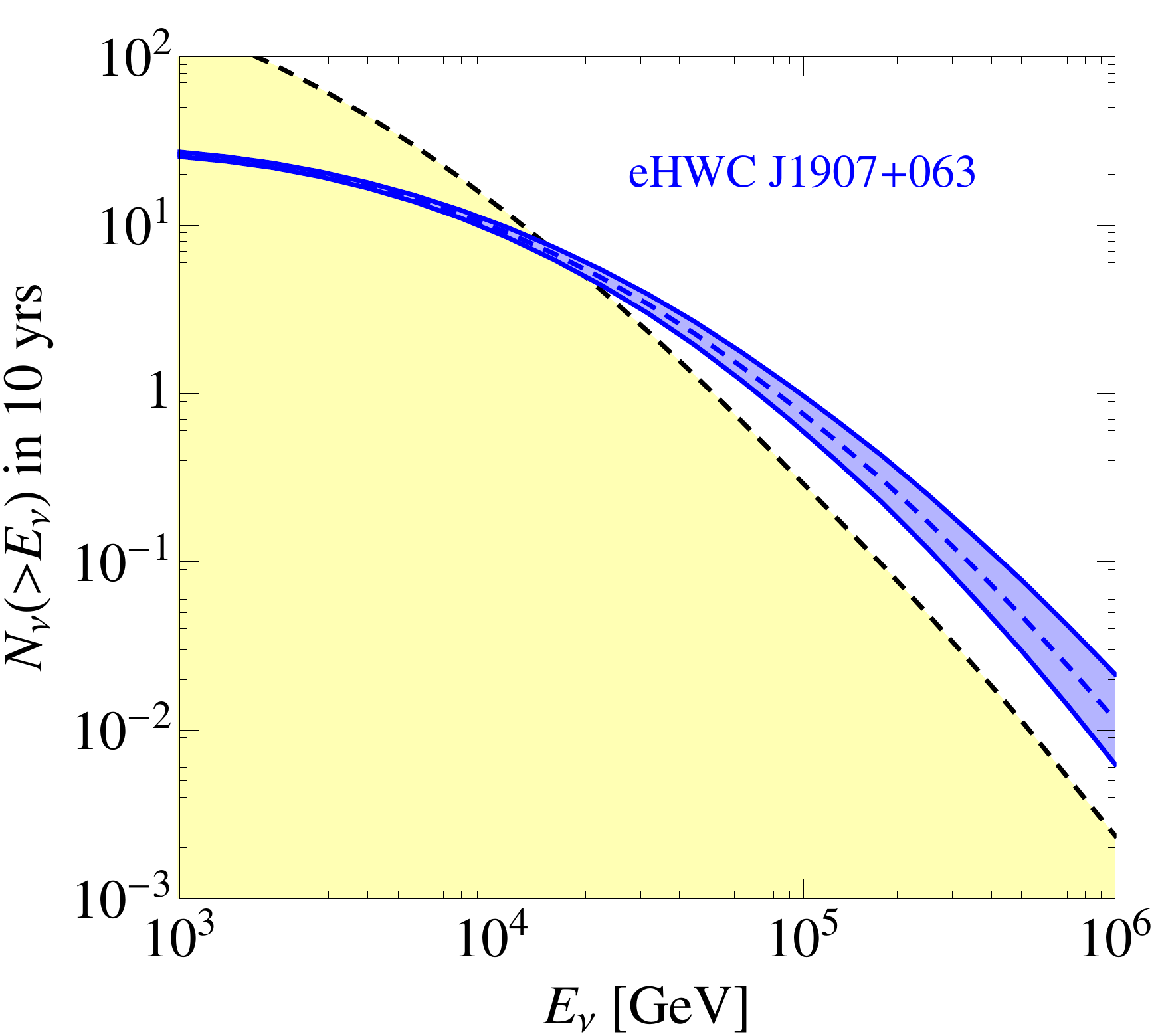} 
\end{tabular}
\caption{\label{fig:gamma_flux_first} 
\underline{Left panel:} Gamma-ray flux for eHWC J1907+063 as reported by the eHWC catalogue, where the blue band encodes the 
statistical error in the $\beta$ parameter. The grey dot-dashed line is the best-fit reported by the 
2HWC catalogue. The shaded grey band encodes the 50\% systematic error 
in the normalization of the fluxes, while the statistical error is not reported. Note that to the best-fit 
given by the 2HWC catalogue we have added a cut-off at 300~TeV energy. 
\underline{Right panel:} Events rate expected at the IceCube detector for the gamma-ray eHWC J1907+063 source in 
10 years running time. The shaded yellow band denotes the atmospheric background.}
\end{figure}

\begin{figure}[!h]
\centering
\begin{tabular}{rl}
\includegraphics[width=0.48\textwidth]{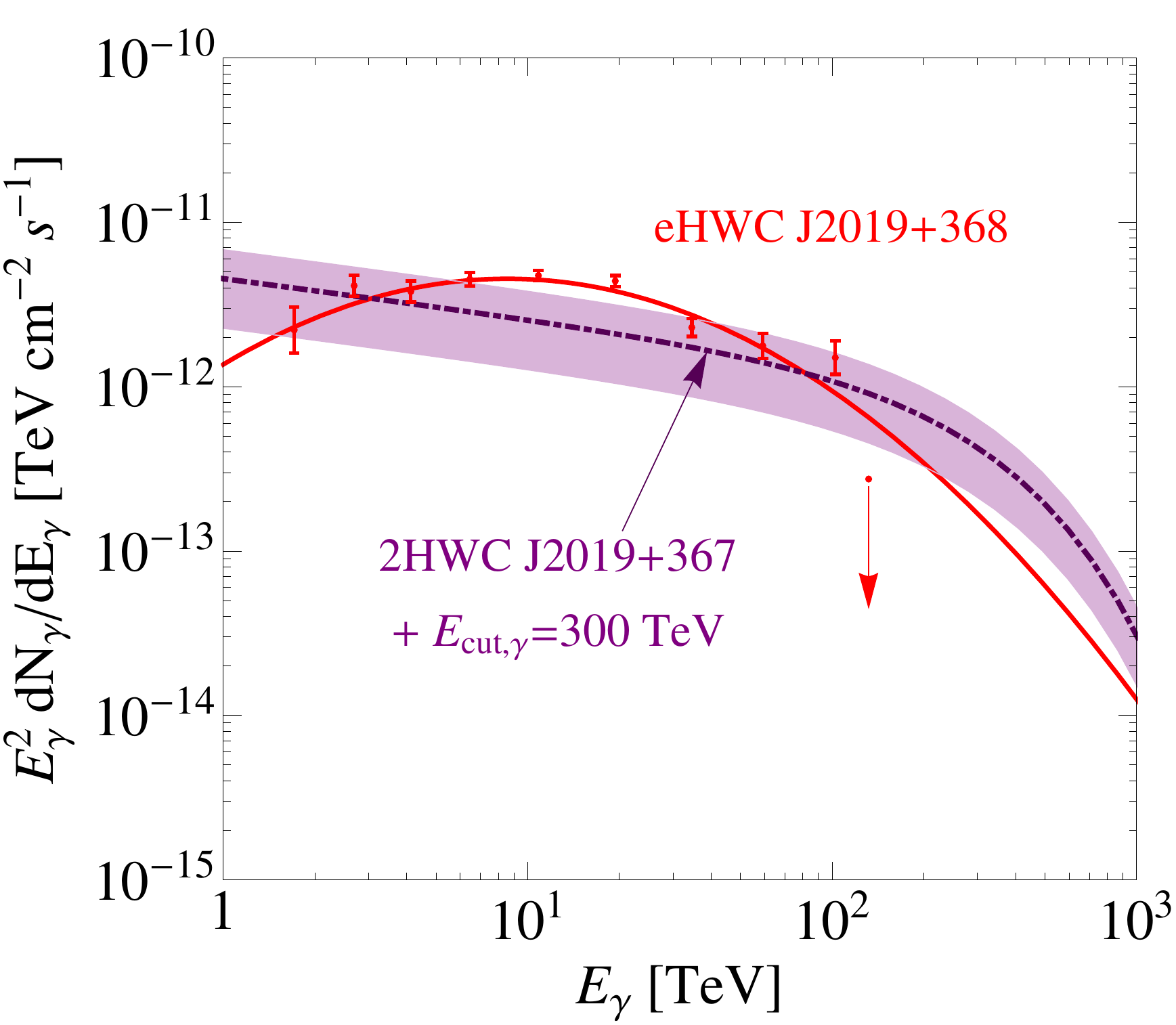} \\
\includegraphics[width=0.48\textwidth]{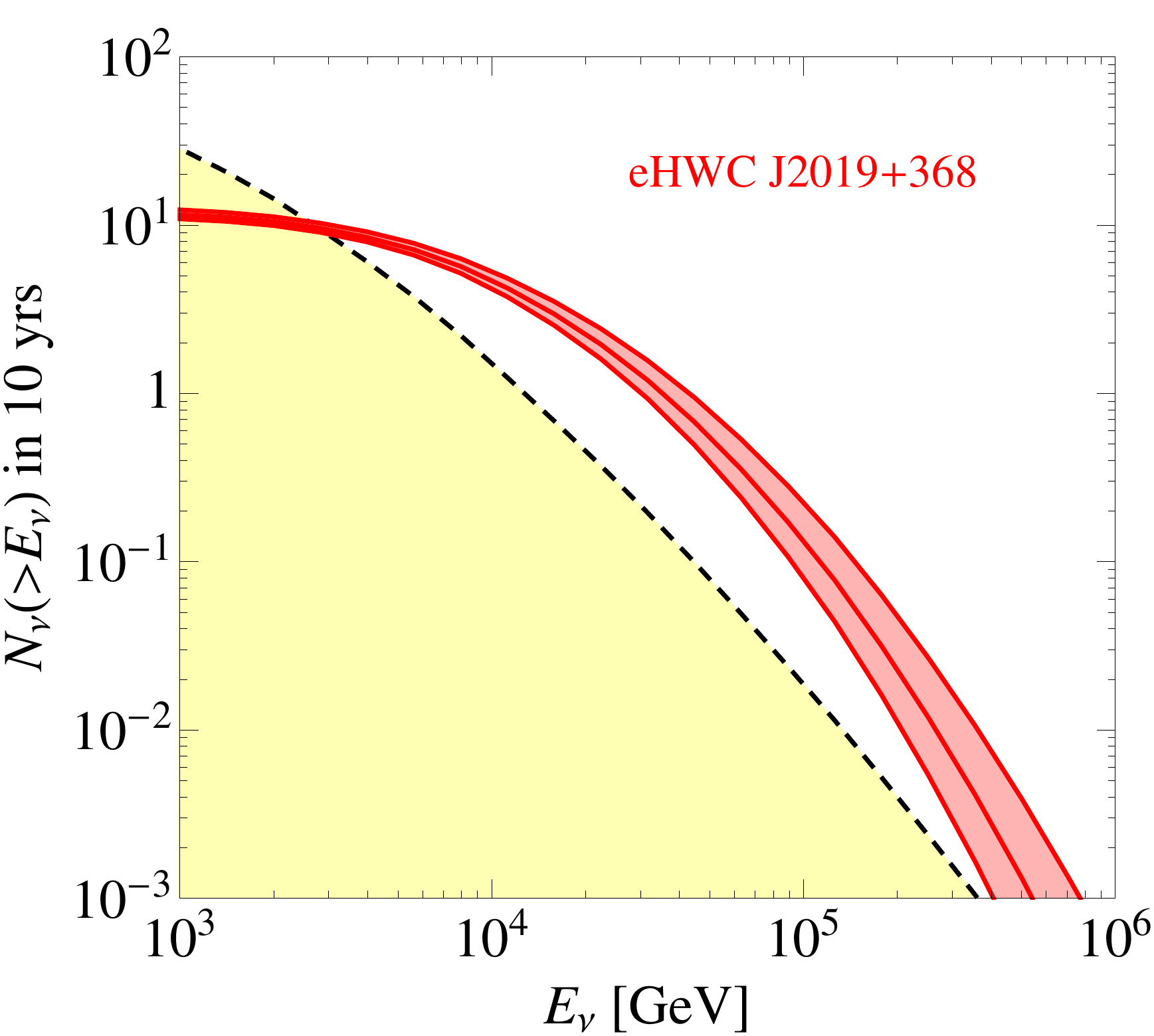}  & 
\includegraphics[width=0.48\textwidth]{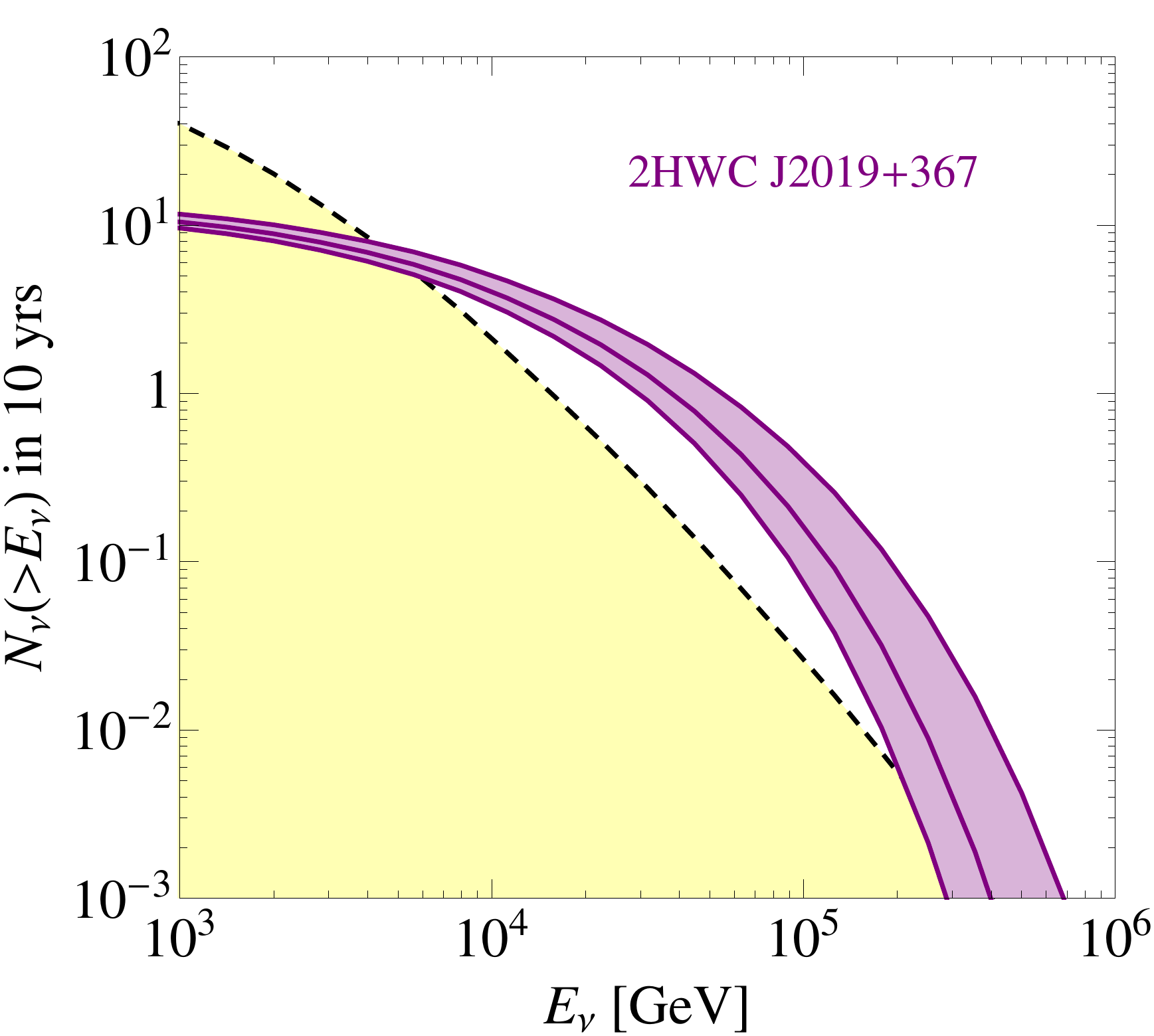}  
\end{tabular}
\caption{\label{fig:gamma_flux_second} 
\underline{Upper panel:} same as Fig.~\ref{fig:gamma_flux_first} but for the eHWC J2019+368 source. 
The red line is the best-fit spectrum reported 
in the eHWC catalogue. The purple dot-dashed line is the best-fit reported by the 
2HWC catalogue. The shaded purple band encodes the 50\% systematic error 
in the normalization of the fluxes, while the statistical error is not reported. 
\underline{Lower panel:} Events rate expected at the IceCube detector for the gamma-ray 
parametrization reported for the eHWC J2019+368 and 2HWC J2019+367 source in 
10 years running time. The red band encodes the statistical error in the $\beta$ parameter, 
while the purple band the 
variation in the cut-off energy parameter $E_{cut, \gamma}$ = 100, 150 and 300~TeV,
The shaded yellow band denotes the atmospheric background. }
\end{figure}

\begin{figure}[!h]
\centering
\begin{tabular}{rl}
\includegraphics[width=0.48\textwidth]{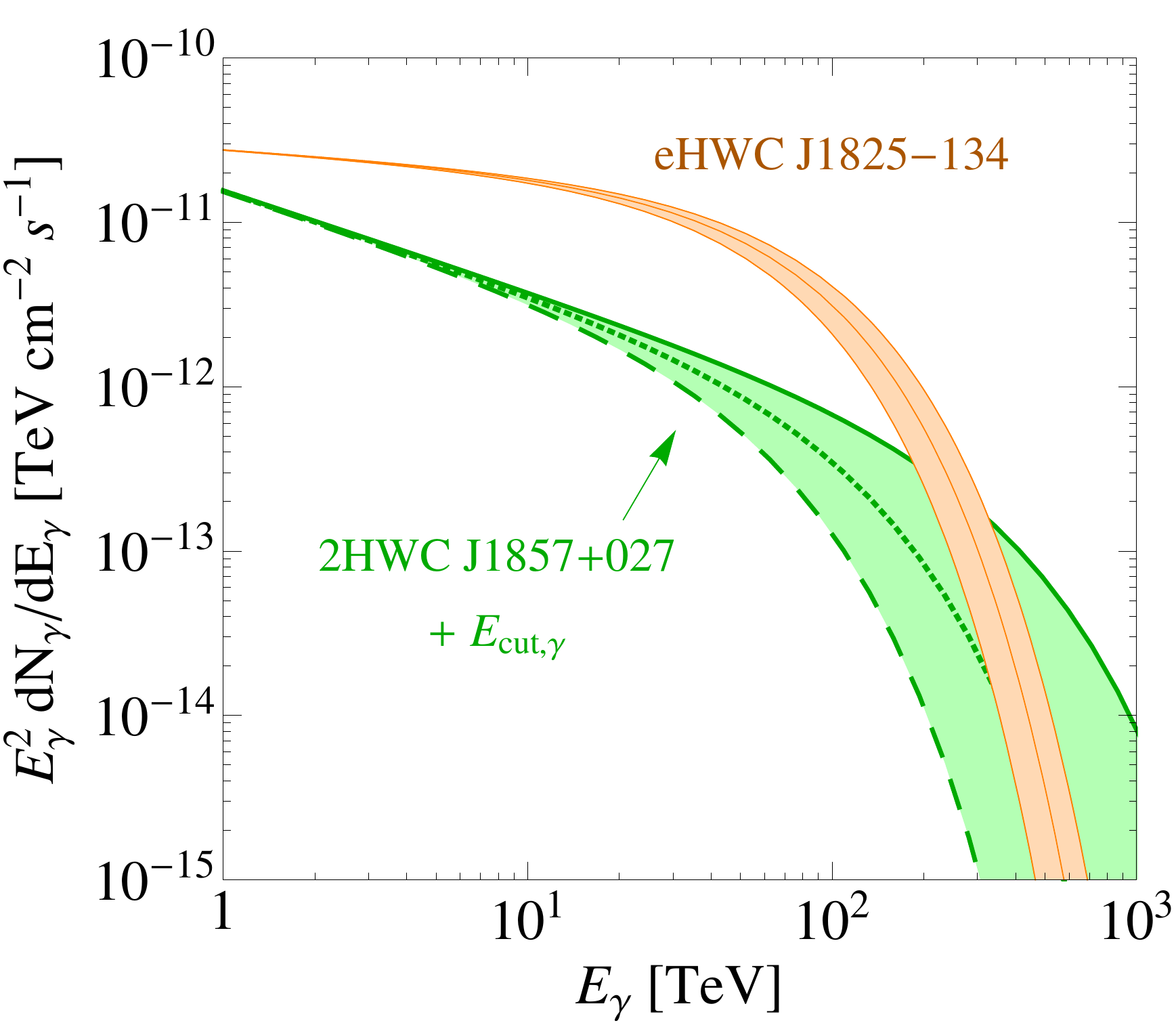} &
\includegraphics[width=0.48\textwidth]{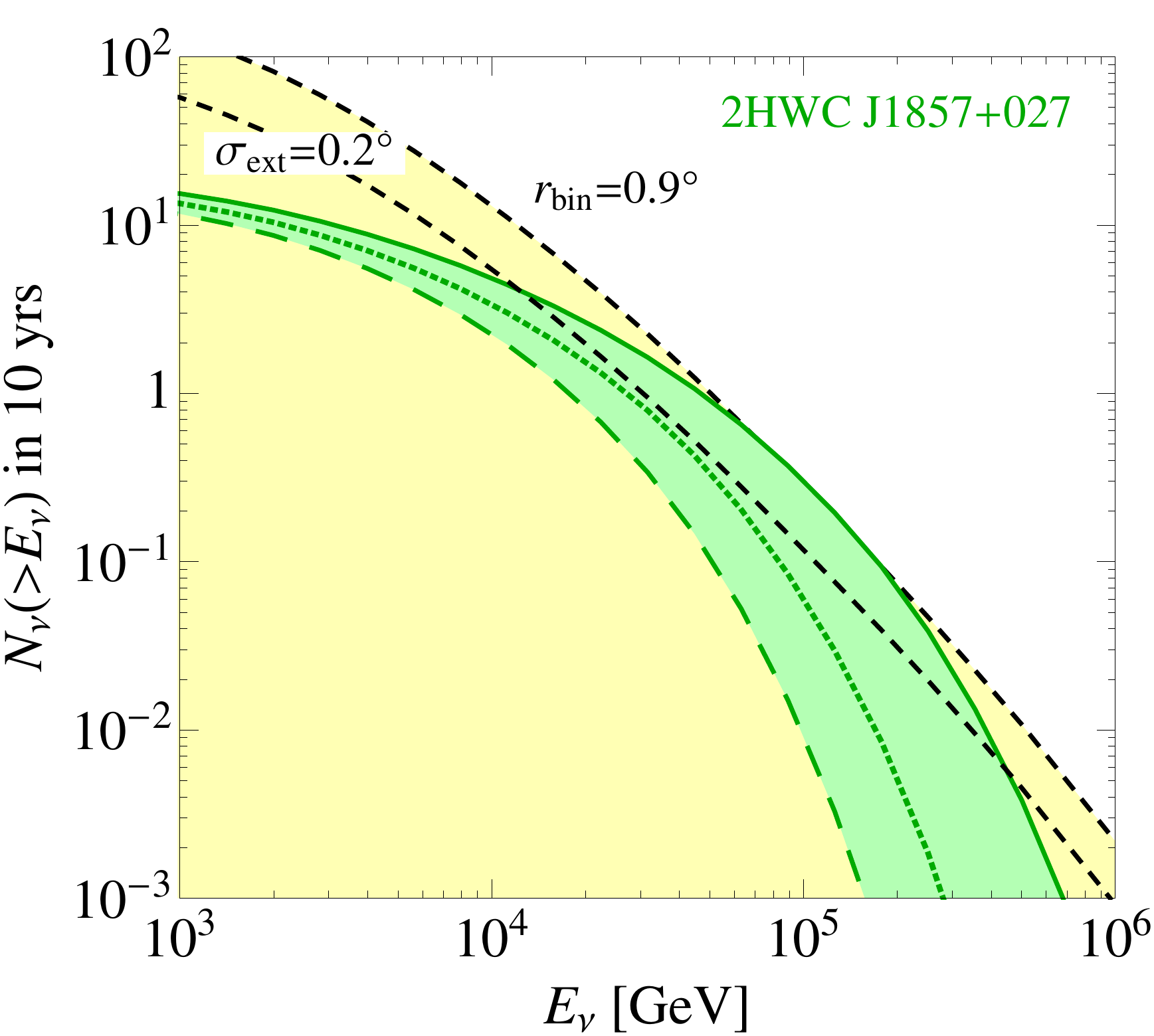}  
\end{tabular}
\caption{\label{fig:gamma_flux_third} 
\underline{Left panel:} Best-fit spectrum for the 2HWC J1857+027 source. The green dashed, dotted and 
solid lines represents different values for $E_{cut, \gamma}$ = 50, 100 and 300~TeV, respectively. 
As comparison, we report also the spectrum for the eHWC J1825-134 source, the most luminous 
source in the eHWC catalogue, with orange lines, where the 
band encodes the statistical error in the parameter $E_{cut, \gamma}$, see text for more details. 
\underline{Right panel:} Events rate expected at the IceCube detector for the gamma-ray 2HWC J1857+027 source in 
10 years running time. The shaded yellow band denotes the atmospheric background. }
\end{figure}

In this work we will consider the possible detection of the eHWC J1907+063, eHWC J2019+368 (2HWC J2019+367) and 
2HWC J1857+027 sources at the 
IceCube detector through tracks events originated by muon neutrino charged current interactions. 

For the effective area of the IceCube detector we use the one reported in 
Ref.~\cite{Stettner:2019tok}, 
where different bands in the zenith angle $\theta_z$ have been considered: 
$-1.00~\leq~cos\theta_z~\leq~-0.75$, $-0.75~\leq~cos\theta_z~\leq~-0.50$, 
$-0.50~\leq~cos\theta_z~\leq~-0.25$ and $-0.25~\leq~cos\theta_z~\leq~0.08$.

The number of events at IceCube can be described by the following expression:
\begin{eqnarray}
N_{\rm ev}= \epsilon_\theta \,t\, \int_{E_\nu^{\rm th}} dE_\nu ~\frac{dN_\nu(E_\nu)}{dE_\nu} \times A_\nu^{\rm eff}(E_\nu,\cos \theta_Z)\,, 
\label{eq:nevmus}
\end{eqnarray}
where a sum over neutrino and antineutrino contributions is implicit. 
The parameter 
$\epsilon_\theta=0.72$ is a reduction factor present because only a fraction of the 
signal will be detected if the source morphology is assumed to be a Gaussian of standard deviation $\sigma_{ext}$. 
For the IceCube detector we have considered $\sigma_{\rm res} \sim 0.4^{\circ}$~\cite{Aartsen:2015mta}. 
The number of neutrino events $\frac{dN_\nu(E_\nu)}{dE_\nu}$ has been calculated 
considering the expressions given in Ref.~\cite{Villante:2008qg}. 
For the calculation of the atmospheric muon neutrinos, following Ref.~\cite{Gonzalez-Garcia:2013iha}, 
we have used the values reported in Refs.~\cite{Honda:2011nf,Volkova:1980sw,Gondolo:1995fq}. 

We estimate the neutrino flux from the eHWC sources as described in Eq.~\ref{eq:nevmus}. 
The results are reported in the right panel of Fig.~\ref{fig:gamma_flux_first} for the 
eHWC J1907+063 source. 
We have fixed the normalization and the size of the source to its best-fit values. 
The spectral index $\alpha_\gamma$ has been also fixed to the best-fit values, while $\beta$ has been varied 
within the statistical errors. 
The results are reported in the lower panels of Fig.~\ref{fig:gamma_flux_second} for the 
eHWC J2019+368 (2HWC J2019+367) source. We have considered the spectral index $\alpha_\gamma$ fixed to its 
best-fit value, as well as the normalization, while the parameter $\beta$ has been varied 
within the statistical errors. Considering the 2HWC parametrization, 
as exemplification, we have fixed the normalization 
to the best-fit reported in the 2HWC catalogue. 
We have then considered an energy cut-off of 100, 150 and 300~TeV and an extension equal to 
the circular bin reported in the 2HWC catalogue ($r_{bin}=0.7^\circ$). 
Finally in the right panel of Fig.~\ref{fig:gamma_flux_third} we report the results for the 
2HWC J1857+027 source. In this case we have considered the best-fit normalization and 
the best-fit value for $\alpha_\gamma$ 
reported in the 2HWC catalogue. Moreover, since the information on the specific morphology of the 
region is currently not public, we have decided to considering also the case in which the emission is gaussian 
with $\sigma_{ext}=0.2^\circ$, and with the same flux as the one given in the 2HWC catalogue.

\section{\label{sec:res} Statistical significance}

The eHWC J1907+063 has been detected by HAWC with an higher flux respect to previous data 
from Atmospheric Cherenkov Telescopes (ACTs) experiments, like HESS. The normalization by HAWC is more compatible 
with the Milagro best-fit~\cite{Abdo:2012jg}. This could be due to the different fields of view 
of the different experiments. Note that in this respect the 
LHAASO~\cite{Bai:2019khm} experiment, that is already running and taking data, 
might give important complementary information on the high-energy tail 
of this source. 
Indeed, the experiment will be able of giving information of sources in the PeV domain. 
Also future experiments, like the planned CTA, 
that will have a sensitivity from 20 GeV up to beyond 300 TeV~\cite{Maier:2019afm}, could 
provide important informations in this respect. 

\begin{figure}[!t]
\centering
\begin{tabular}{rl}
\includegraphics[width=0.48\textwidth]{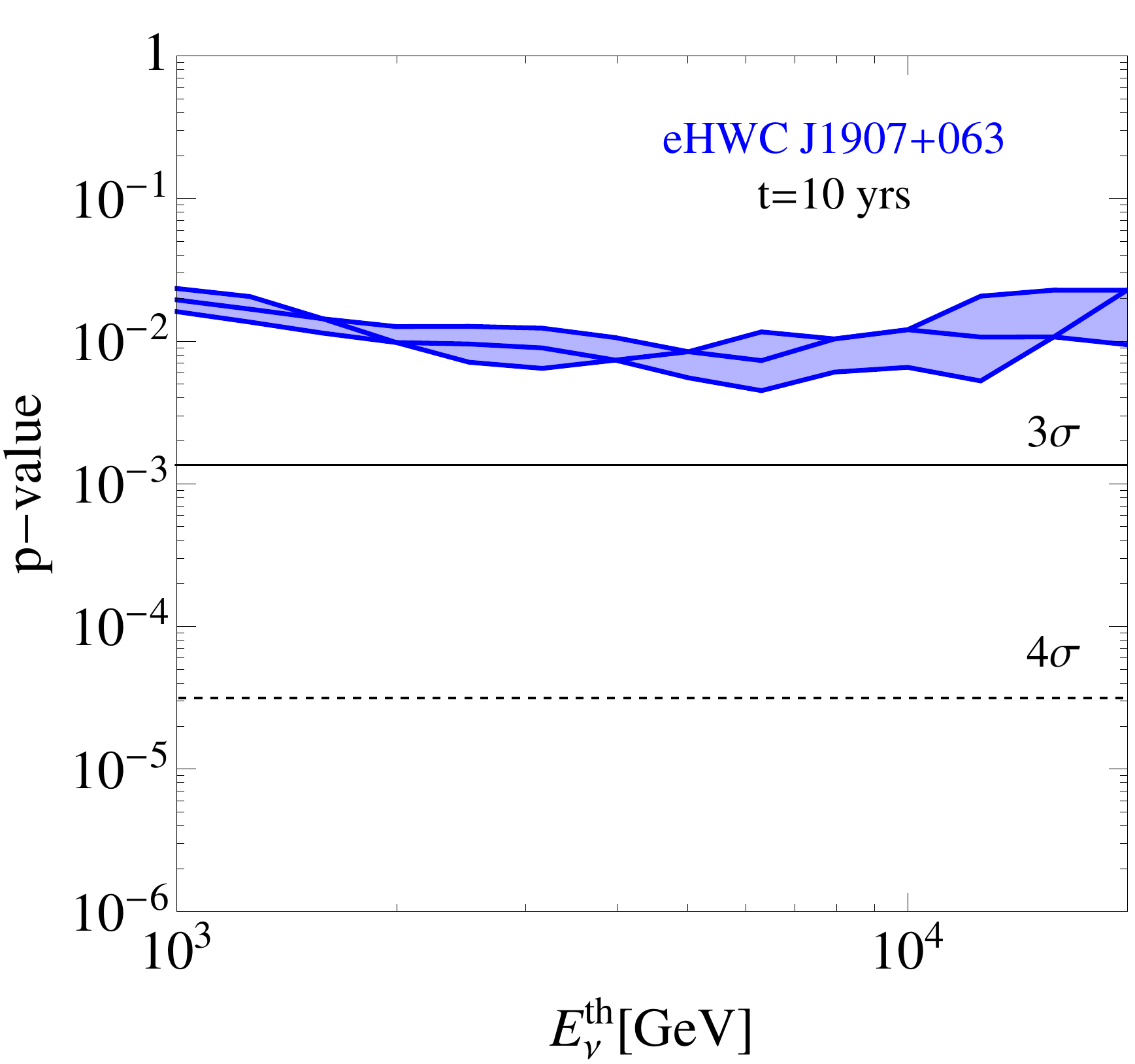} & 
\includegraphics[width=0.48\textwidth]{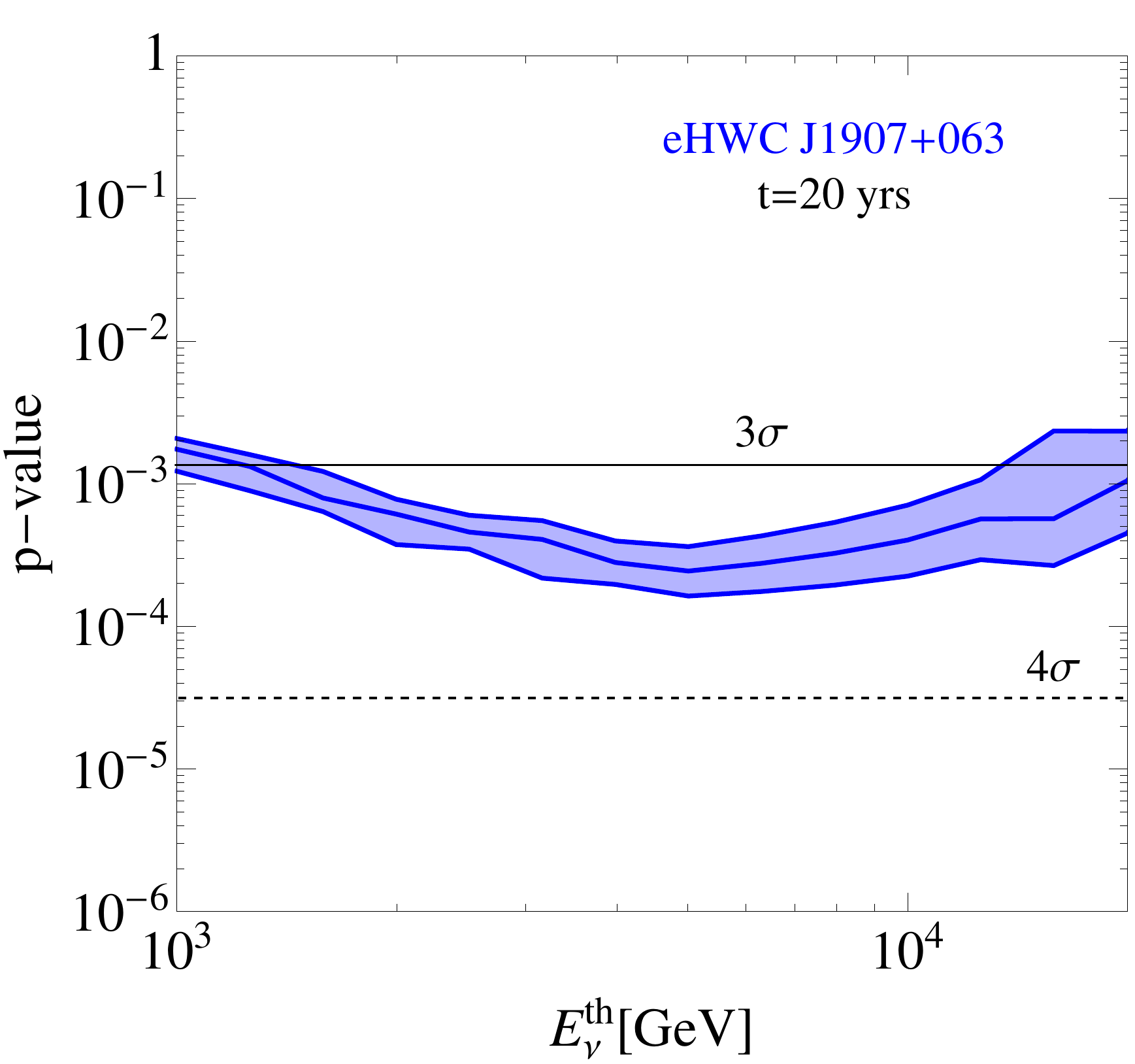}
\end{tabular}
\caption{\label{fig:p_value_first} 
\underline{Left panel:} Statistical significance expected at the IceCube detector 
for the eHWC J1907+063 source after 10 years running time. 
\underline{Right panel:} Same as the left panel but for 20 years running time.}
\end{figure}

We have estimated the statistical significance as reported in Ref.~\cite{ATLAS:2011tau} 
and as described in Refs.~\cite{Halzen:2016seh,Gonzalez-Garcia:2013iha}. 
We report in Fig.~\ref{fig:p_value_first} the results for the p-value as a function of the energy threshold for 
10 and 20 years of running time of the IceCube detector. 
We find for the eHWC J1907+063 source a p-value of about 1\% in 10 years, almost independently of the energy threshold 
used in the analysis, while in 20 years a 3$\sigma$ is reached for an energy threshold of 
about~1~TeV. The IceCube detector currently has reported a p-value of about 1\% from this source~\cite{Aartsen:2018ywr}. 
Note, that the IceCube point source analysis uses an unbinned likelihood method, that takes into account 
the energy distribution of the events with their individual angular uncertainties. 
This source, previously identified as MGRO J1908+06, was already considered as one of the most promising source 
to be detected by IceCube, because neutrinos from this source can reach the detector without significant 
absorption in the Earth~\cite{Halzen:2008zj,Kappes:2009zza,GonzalezGarcia:2009jc,Gonzalez-Garcia:2013iha,Halzen:2016seh}. 

\begin{figure}[!t]
\centering
\begin{tabular}{rl}
\includegraphics[width=0.48\textwidth]{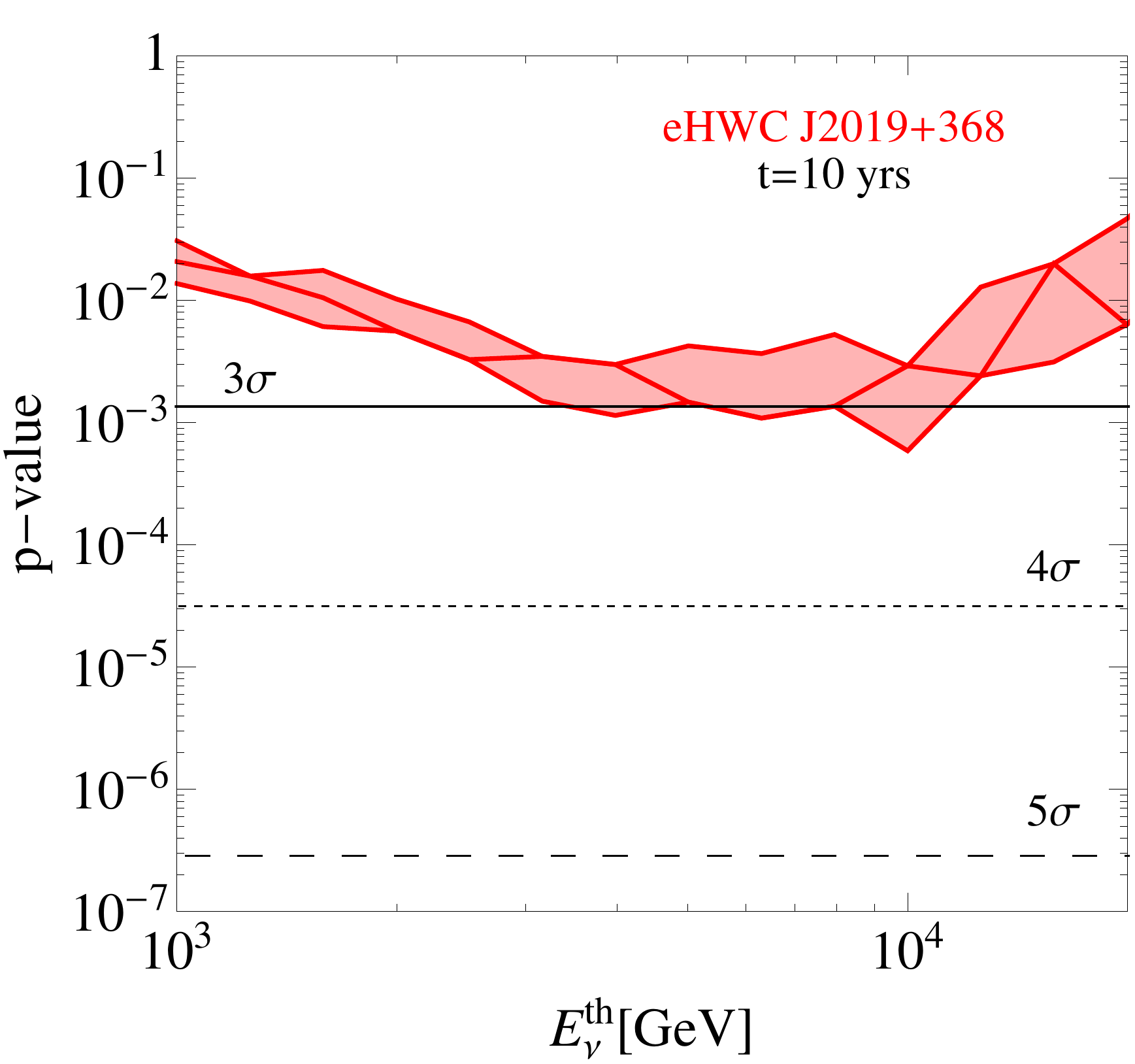} &
\includegraphics[width=0.48\textwidth]{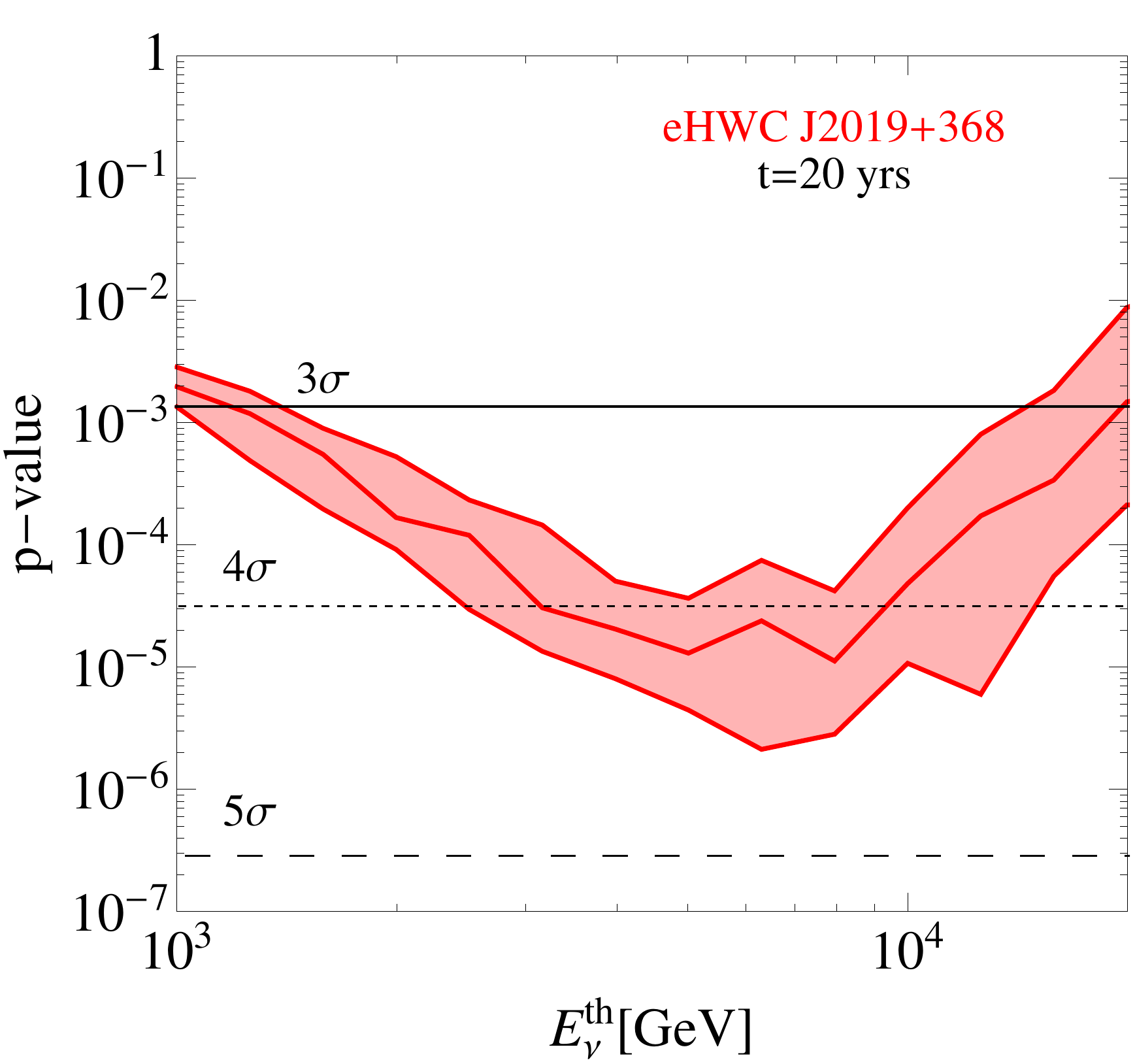}
\end{tabular}
\caption{\label{fig:p_value_second_red} 
\underline{Left panel:} Statistical significance expected at the IceCube detector 
for the eHWC J2019+368 source after 10 years running time. 
\underline{Right panel:} Same as the left panel but for 20 years running time.}
\end{figure}

\begin{figure}[!t]
\centering
\begin{tabular}{rl}
\includegraphics[width=0.48\textwidth]{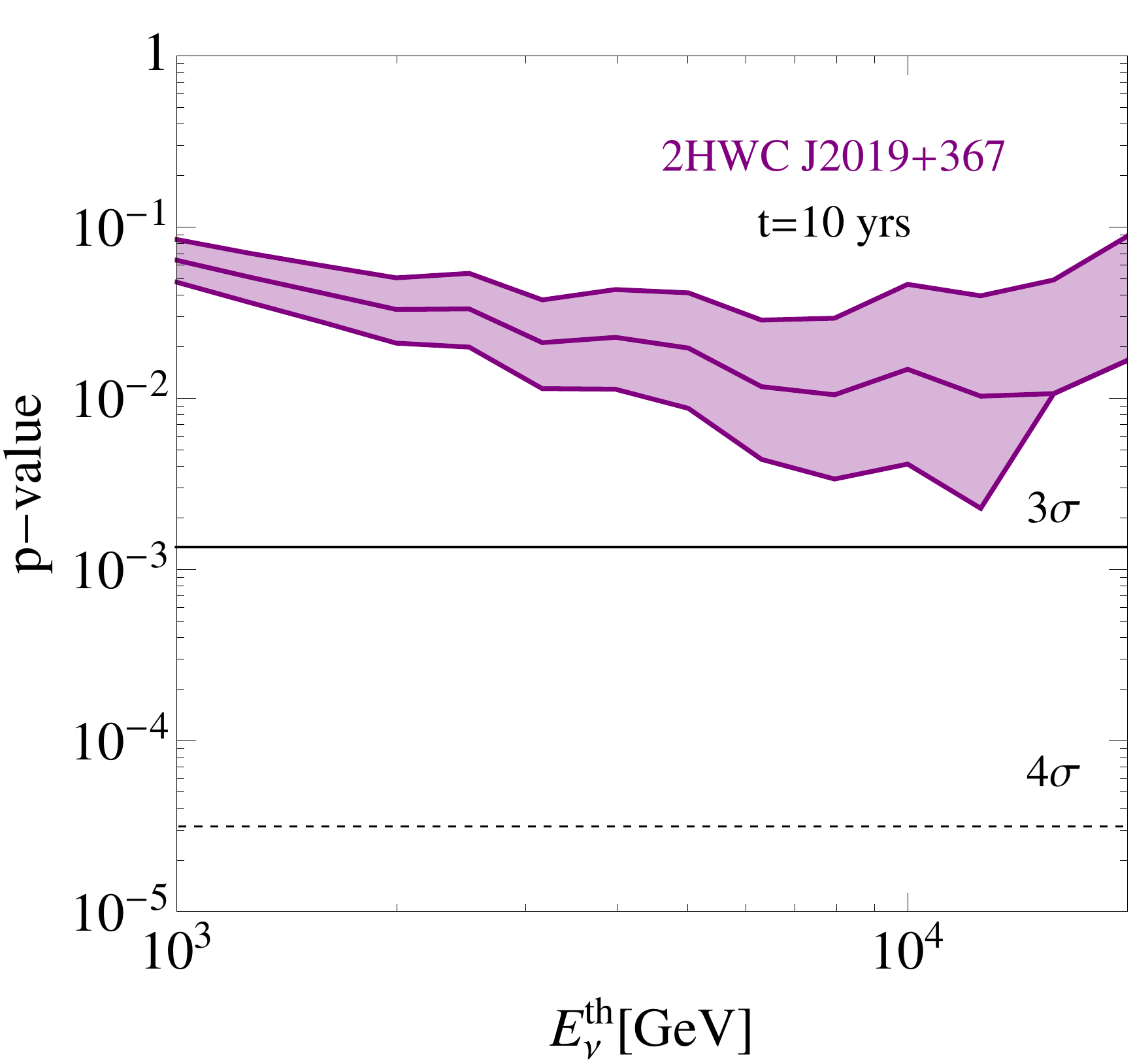} &
\includegraphics[width=0.48\textwidth]{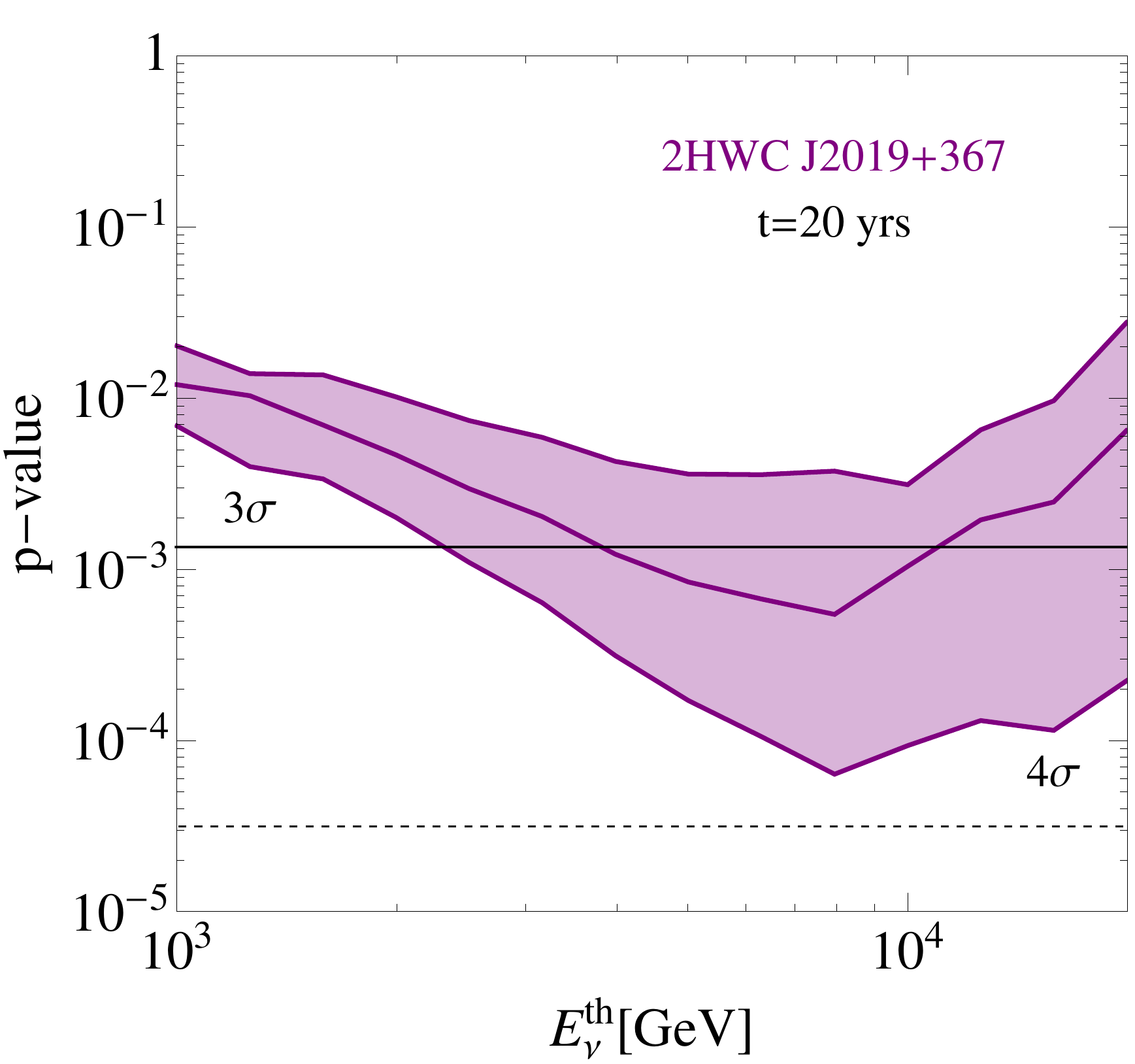}
\end{tabular}
\caption{\label{fig:p_value_second} 
\underline{Left panel:} Statistical significance expected at the IceCube detector 
for the 2HWC J2019+367 source after 10 years running time. 
\underline{Right panel:} Same as the left panel but for 20 years running time.}
\end{figure}

The second most luminous source, eHWC J2019+368, belongs to the Cygnus region, that is a star-forming region. 
A more complex picture for this source might be present since the physics behind the production might be more complex, see 
for example~\cite{Gabici:2007qb} about the production of neutrinos in association with supernova remnants and 
molecular clouds, and Ref.~\cite{Yoast-Hull:2017gaj} for an analysis of the Cygnus region, where it 
was shown that a detection from the whole Cygnus region is probable with IceCube. 
We report in Fig.~\ref{fig:p_value_second_red} the results for the p-value as a function of the energy threshold for 
10 and 20 years of running time of the IceCube detector for the eHWC J2019+368 parametrization and 
extension, while in Fig.~\ref{fig:p_value_second} the result for the 2HWC J2019+367 region, assuming 
a tested radius $r_{bin}=0.7^\circ$. 
As can be seen from the figures, considering the eHWC J2019+368 parametrization and 
extension, a detection with a p-value of about 1\% could be reach in 
10 years ($3\sigma$ for an energy threshold of about 10~TeV), while it could reach about $3\sigma$ or more 
($4\sigma$ for an energy threshold of about 10~TeV) in 20 years running time. 
Considering, instead, the 2HWC J2019+367 parametrization and extension ($r_{bin}$=0.7$^\circ$), 
almost 3$\sigma$ could be reached in  
10 years for an energy threshold of 10~TeV in neutrinos, while almost $4\sigma$ could be reach in 20 years, 
if the cut-off energy is of about 300~TeV. 
Note that, even if the detection of neutrinos from the 2HWC J2019+367 region might 
be difficult, the discrimination between the search in a smaller region, the 
eHWC J2019+368 parametrization and 
extension, or from the full 
2HWC J2019+367 region could be important also to discriminate between different production mechanism, i.e. 
if for example the neutrinos are produced by the single source or by a more complex picture of supernova 
remnant and molecular clouds. 
Note that in a previous analysis~\cite{Halzen:2016seh} it was found that we expect to obtain about 3$\sigma$ discovery in roughly 15 years at 
IceCube, considering the spectrum reported by VERITAS. This is consistent with what reported for 
the eHWC J2019+368 parametrization and extension. 

\begin{figure}[!t]
\centering
\begin{tabular}{rl}
\includegraphics[width=0.48\textwidth]{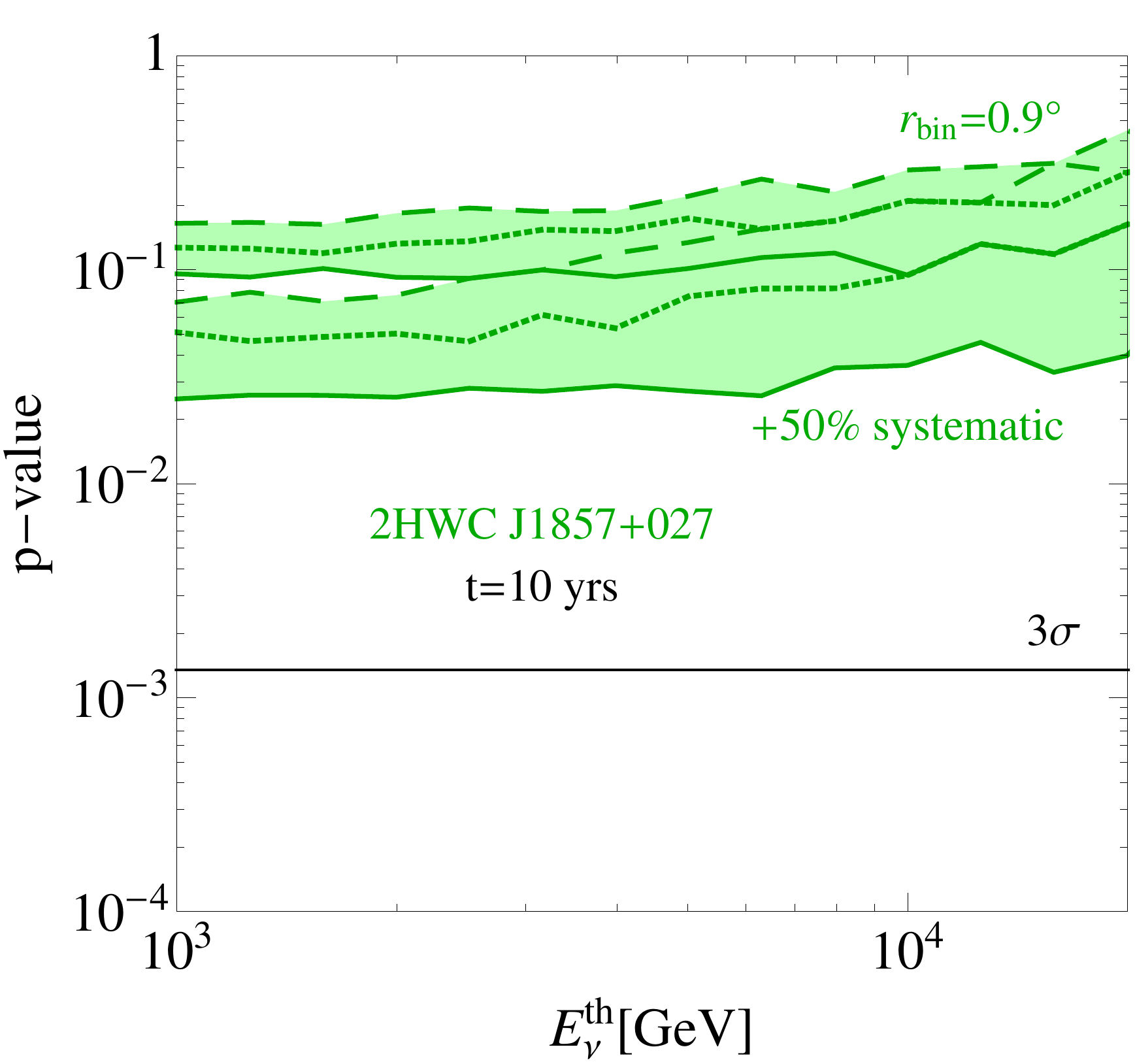} &
\includegraphics[width=0.48\textwidth]{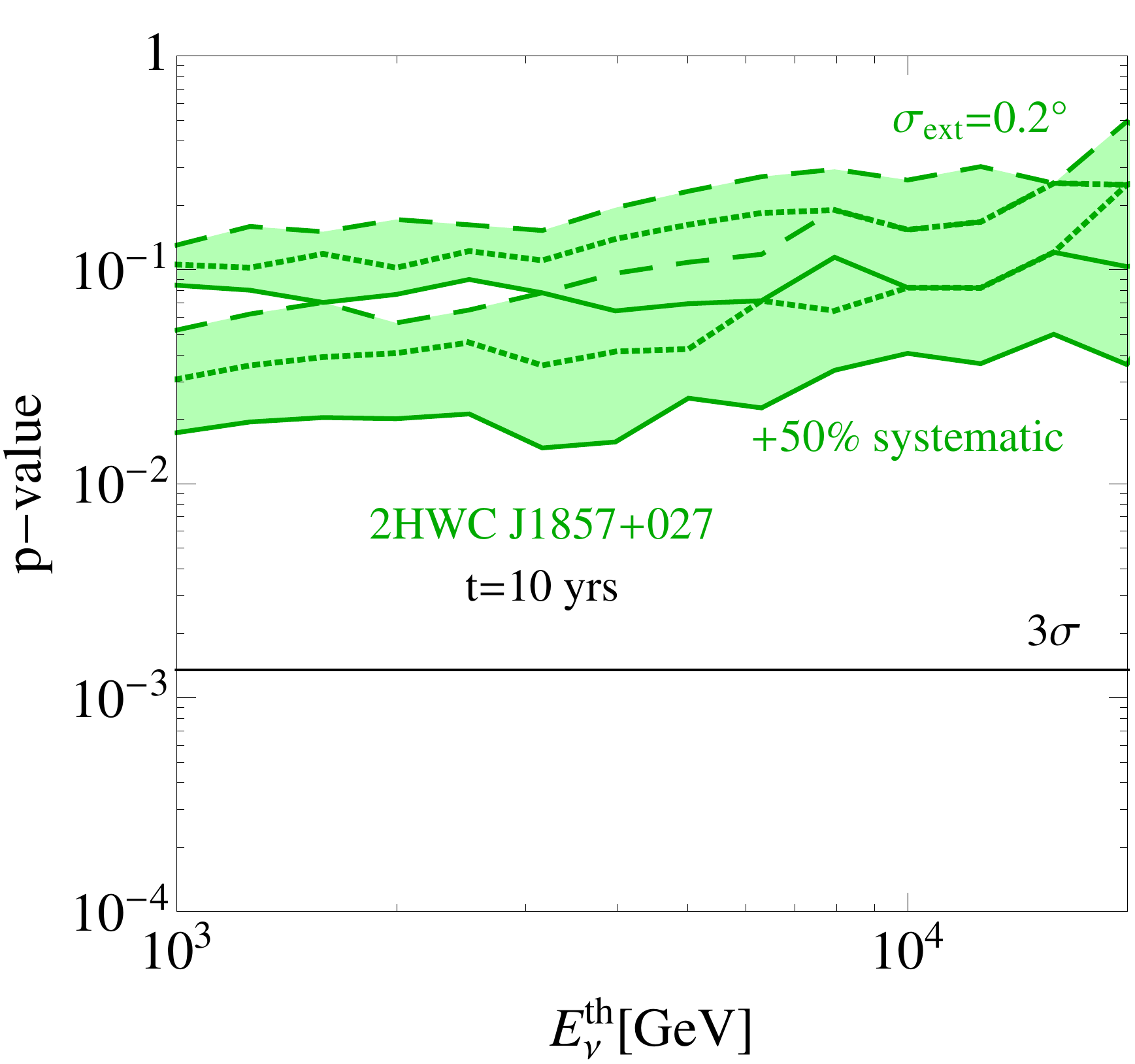}
\end{tabular}
\caption{\label{fig:p_value_third} 
\underline{Left panel:} Statistical significance expected at the IceCube detector 
for the 2HWC J1857+027 source after 10 years running time. 
\underline{Right panel:} Same as the left panel but for a gaussian morphology with 
$\sigma_{ext}=0.2^\circ$.}
\end{figure}

\begin{figure}[!t]
\centering
\begin{tabular}{rl}
\includegraphics[width=0.48\textwidth]{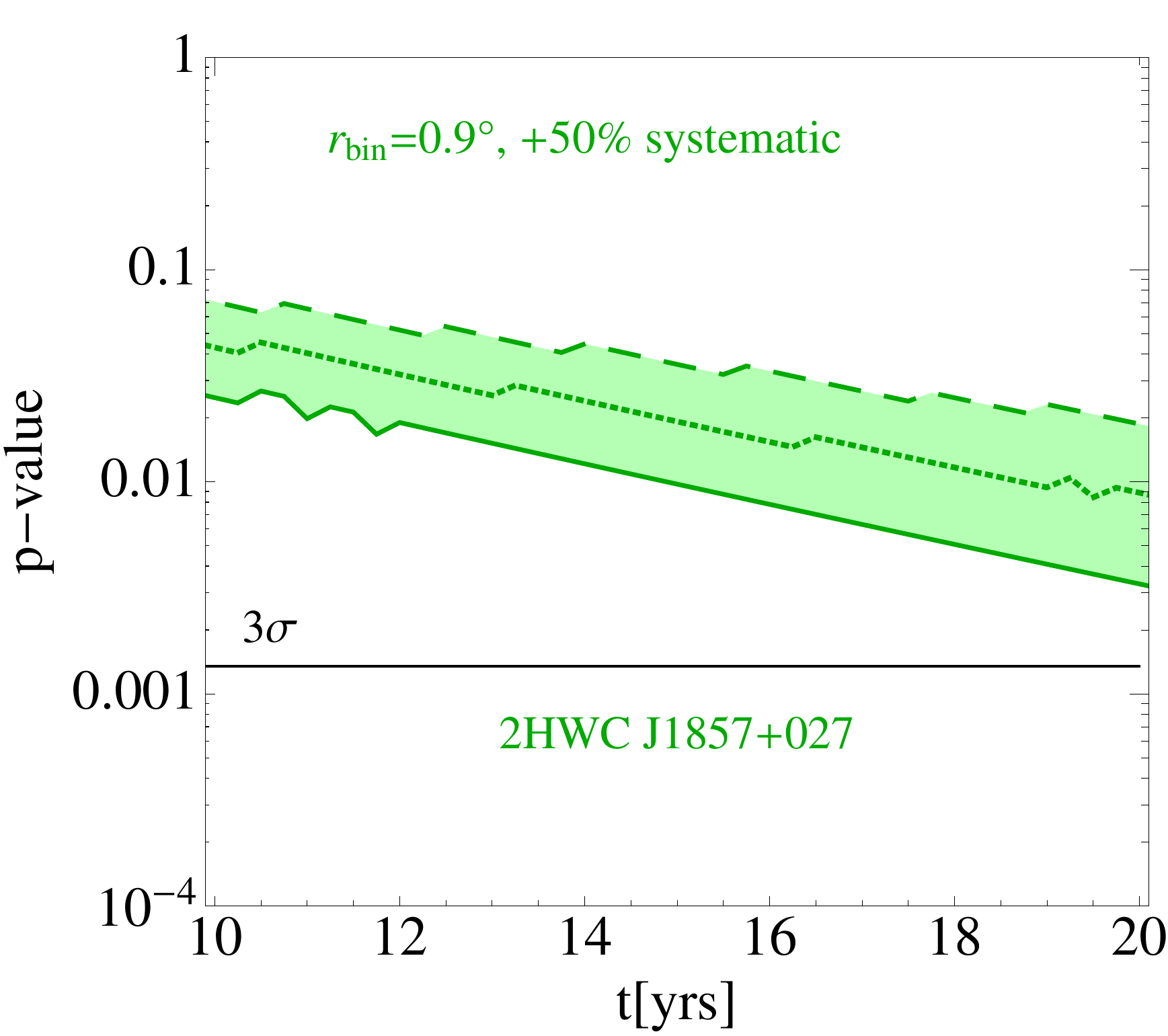} &
\includegraphics[width=0.48\textwidth]{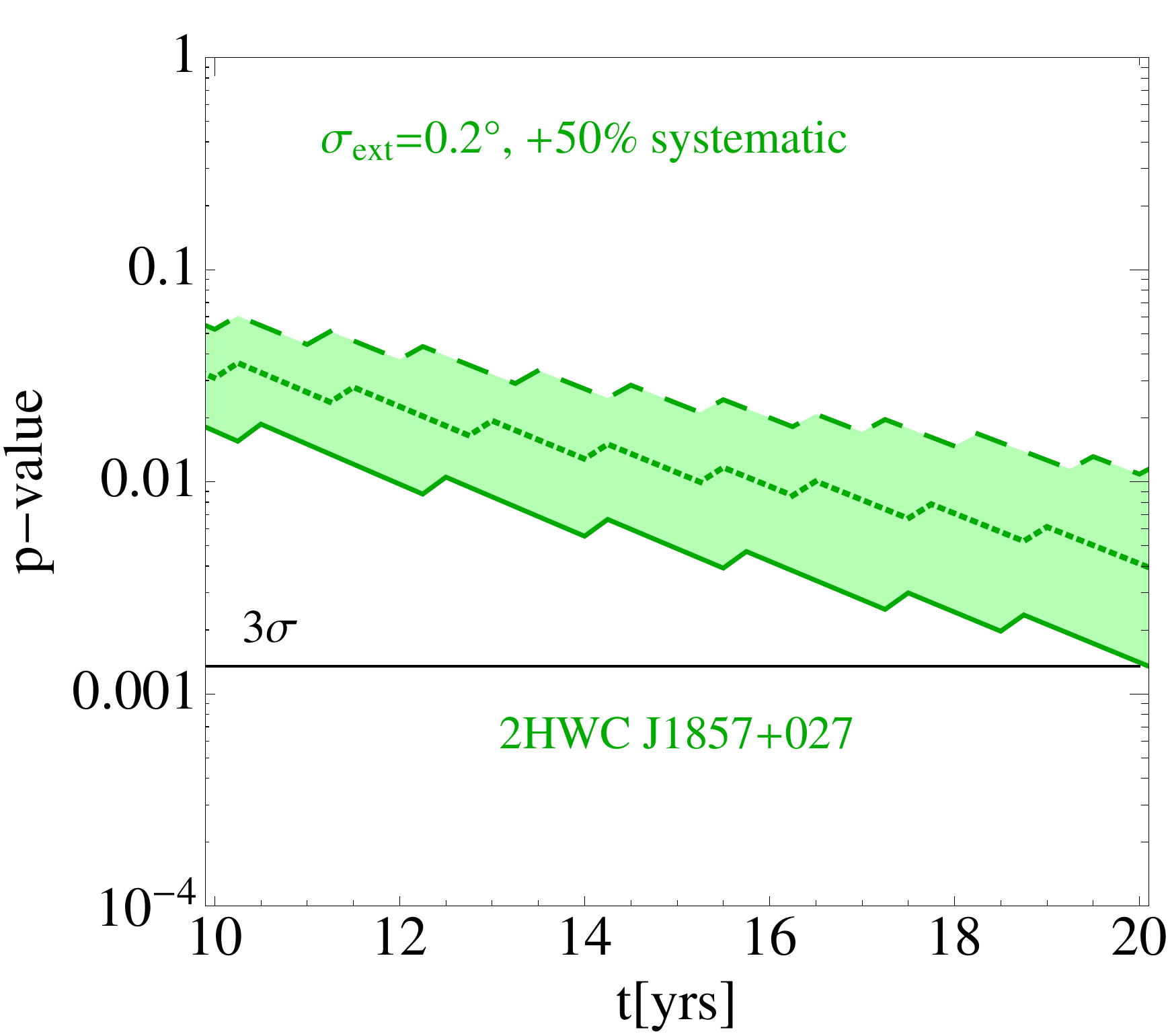}
\end{tabular}
\caption{\label{fig:p_value_third_time} Dependence of the statistical significance expected at the IceCube detector 
for the 2HWC J1857+027 source on the running time. }
\end{figure}

Finally, we report in Fig.~\ref{fig:p_value_third} the results for the p-value as a function of the energy threshold for 
10 years of running time of the IceCube detector for the 2HWC J1857+027 source, while in 
Fig.~\ref{fig:p_value_third_time} it is shown explicitly the dependence on the running time. 
In this case, we have considered also the systematic error in the normalization of the flux and 
a cut-off energy of 50, 100, 300~TeV, having however in mind that 
the latter value is in tension with the fact that the source has not been reported in the 
eHWC catalogue. 
Moreover, in the 2HWC catalogue a specific circular bin has been considered for the 
search, giving however no explicit information on the morphology of the source. 
For this reason, we have considered the circular bin reported in the 2HWC catalogue, 0.9$^\circ$, 
as well as a gaussian morphology with extension of 0.2$^\circ$. 
The IceCube detector has recently reported a p-value of about 2\% from this region. Within our 
statistical method, 
we find that this could be possible only considering the systematic error on the normalization 
of the flux. Moreover, a better agreement is found considering a gaussian morphology with extension of about 0.2$^\circ$ for the 
source and an energy cut-off greater than 100~TeV. 
Since this source has not been detected in the eHWC catalogue, this 
represents a puzzling results that needs additional data to clarify the situation. 
This conclusion indicates that 
we are entering the era of precision physics both on multi TeV gamma-ray and on high-energy neutrino astronomy. 
For this reason a synergy between these two types of experiments, gamma-ray astronomy and neutrinos, will be 
important to shed light on the origin of galactic cosmic-rays and on the characteristics of the source.


\section{\label{sec:conclusion} Conclusions }

Using updated information on the spectrum provided by the 
HAWC collaboration, we have calculated 
the number of events expected at the IceCube detectors for 
the two brightest sources eHWC J1907+063 and eHWC J2019+368. 
For the latter we have considered also how things change considering 
the extended region 2HWC J2019+367. 
Since an excess in neutrinos is present from 2HWC J1857+027, we have 
also studied the neutrino emission from this source. 

Moreover, 
we have calculated the statistical significance for these sources 
at the IceCube detector considering 10 and 20 years running time. 
We found that more than 3$\sigma$ can be reached in the next decade, independently of the 
statistical errors, for the sources eHWC J1907+063 and eHWC J2019+368. 
Considering the 2HWC J2019+367 region, instead, a detection at about 3$\sigma$ or more is 
expected in 20 years of running time, for a neutrino energy threshold of about 10~TeV. 
For the source 2HWC J1857+027 we have explicitly shown the dependence on the 
systematic error of the normalization, on the cut-off 
energy and on the extension of the source for the p-value. Assuming a gaussian morphology, 
an high value of the normalization (within the 
+50\% error), an extension of 0.2$^\circ$ or smaller 
and a cut-off energy of 100~TeV or larger, we find a p-value of a few percent in 
10 years. Consistency checks with the fact that 
this source does not belong to the eHWC catalogue are to further be clarified with 
future gamma-ray and neutrino data. 

Respect to previous works we have used updated information on the spectra from 
the eHWC and 2HWC catalogue and we have also added the source 2HWC J1857+027 to the analysis, showing explicitly 
the dependence on the systematic error of the normalization, on the energy cut-off and on the extension of the source. 

The possibility of detecting these sources with the future 
KM3NeT detector depends on the visibility of these sources at KM3NeT, that is in general below or 
close to 50\%. Considering a latitude of of 36º 16' N for the KM3NeT detector and the 
expression of the visibility as reported in Ref.~\cite{Costantini:2004ap}, we found that 
$\epsilon_v=0.47, 0.31, 0.49$ for eHWC J1907+063, eHWC J2019+368, 2HWC~J1857+027, respectively. 
Note also that the visibility can increase considering 6 or 10 degrees above the horizon, 
thus increasing the sensitivity to these sources, see~\cite{Adrian-Martinez:2016fdl, 
Aiello:2018usb}. 
The most promising sources to be detected at KM3NeT are eHWC J1907+063 and 
2HWC J1857+027. For the 
source in the Cygnus region, its position is not optimal for a detector in the northern hemisphere. 
Similar considerations holds true for the Baikal-GVD detector~\cite{Avrorin:2015skm}, for which we considered a 
latitude of 51º N and the visibilities that we found are $\epsilon_v=0.46, 0.13$ and $0.48$. 

We want here to comment on the possibility of detecting these sources with IceCube Gen2~\cite{Aartsen:2014njl,Aartsen:2020fgd}. 
IceCube Gen2 is planned to have an effective area of about five times bigger than the current 
IceCube detector, see 
Ref.~\cite{Aartsen:2014njl}. Thus, even one year of running of this bigger detector could 
improve the sensitivity to these sources dramatically.

\section*{Acknowledgements}
This project has received funding from the European Union's Horizon 2020 research and innovation programme under 
the Marie Sk\l{}odowska-Curie grant agreement No. 843418. 
VN acknowledges Stefano Gabici, Francis Halzen and 
Ali Kheirandish for discussions on the subject and 
for reading the manuscript. 
VN acknowledges Luigi Fusco for discussions about the atmospheric neutrino flux.

\clearpage

\bibliographystyle{elsarticle-num.bst}
\bibliography{biblio}


\end{document}